\documentclass[sigconf]{acmart}

\usepackage{booktabs} 
\usepackage{subfig}
\usepackage{graphicx}

\usepackage{balance}
\usepackage{tikz}
\usepackage{float}
\usepackage{placeins}
\usetikzlibrary{arrows}
\usepackage{makecell}
\usepackage{bbm}

\definecolor{mygray}{RGB}{50, 50, 50}

\newcommand{\maxf}[1]{\ensuremath{\mathbf{#1}}}
\newcommand{\maxff}[2]{{\fontseries{#1}\selectfont #2}\par}

\copyrightyear{2019} 
\acmYear{2019} 
\setcopyright{acmcopyright}
\acmConference[WSDM '19]{The Twelfth ACM International Conference on Web Search and Data Mining}{February 11--15, 2019}{Melbourne, VIC, Australia}
\acmBooktitle{The Twelfth ACM International Conference on Web Search and Data Mining (WSDM '19), February 11--15, 2019, Melbourne, VIC, Australia}
\acmPrice{15.00}
\acmDOI{10.1145/3289600.3290961}
\acmISBN{978-1-4503-5940-5/19/02}

\ccsdesc[500]{Information systems~Link and co-citation analysis}
\ccsdesc[500]{Human-centered computing~Graph drawings}
\ccsdesc[500]{Computing methodologies~Dimensionality reduction and manifold learning}
\ccsdesc[300]{Computing methodologies~Factorization methods}
\ccsdesc[300]{Computing methodologies~Learning latent representations}

\begin{document}

\title{Spring-Electrical Models For Link Prediction}

\author{Yana Kashinskaya}
\affiliation{
   \institution{Yandex School of Data Analysis\\Moscow, Russia}
}
\email{kashin.yana@gmail.com}

\author{Egor Samosvat}
\authornote{This work is supported by Russian President grant MK-527.2017.1.}
\affiliation{
   \institution{Yandex\\Moscow, Russia}
}
\email{sameg@yandex-team.ru}

\author{Akmal Artikov}
\affiliation{%
  \institution{Yandex\\Moscow, Russia}
}
\email{aartikov@yandex-team.ru}

\begin{abstract}
We propose a link prediction algorithm that is based on spring-electrical models. The idea to study these models came from the fact that spring-electrical models have been successfully used for networks visualization. 
A good network visualization usually implies that nodes similar in terms of network topology, e.g., connected and/or belonging to one cluster, tend to be visualized close to each other. Therefore, we assumed that the Euclidean distance between nodes in the obtained network layout correlates with a probability of a link between them. We evaluate the proposed method against several popular baselines and demonstrate its flexibility by applying it to undirected, directed and bipartite networks. 
\end{abstract}

%
%


\keywords{Spring-electrical models, Link prediction, Graph embeddings}
\maketitle

\section{Introduction} \label{introduction}

Link prediction is usually understood as a problem of predicting missed edges in partially observed networks or predicting edges which will appear in the near future of evolving networks~\cite{menon2011link}. A prediction is based on the currently observed edges and takes into account a topological structure of the network. Also, there may be some side information or meta-data such as node and edge attributes. 

The importance of link prediction problem follows naturally from a variety of its practical applications. For example, popular online social networks such as Facebook and LinkedIn suggest a list of people you may know. Many e-commerce websites have personalized recommendations which can be interpreted as predictions of links in bipartite graphs \cite{schafer1999recommender}. Link prediction can also help in the reconstruction of some partially studied biological networks by allowing researches to focus on the most probable connections~\cite{lu2011link}. 

More formally, in order to evaluate the performance of the proposed link prediction method we consider the following problem formulation. The input is a partially observed graph and our aim is to predict the status (existence or non-existence) of edges for unobserved pairs of nodes. This definition is sometimes called structural link prediction problem~\cite{menon2011link}. Another possible definition suggests to predict future edges based on the past edges but it is limited to time-evolving networks which have several snapshots.

An extensive survey of link prediction methods can be found in~\cite{liben2007link,lu2011link,cukierski2011graph}.
Here we describe some of the most popular approaches that are usually used as baselines for evaluation~\cite{menon2011link,sherkat2015structural}.
The simplest framework of link prediction methods is the similarity-based algorithm, where to each pair of nodes a score is assigned based on topological properties of the graph~\cite{lu2011link}. This score should measure similarity (also called proximity) of any two chosen nodes. For example, one such score is the number of common neighbours that two nodes share, because usually if nodes have a lot of common neighbours they tend to be connected with each other and belong to one cluster. Other popular scores are Shortest Distance, Preferential Attachment~\cite{barabasi1999emergence},  Jaccard~\cite{small1973co} and Adamic-Adar score~\cite{adamic2003friends}.

Another important class of link prediction methods are latent feature models \cite{miller2009nonparametric,menon2010log,menon2011link,dunlavy2011temporal,acar2009link}. The basic idea is to assign each node a vector of latent features in such a way that connected nodes will have similar latent features. Many approaches from this class are based on the matrix factorization technique which gained popularity through its successful application to the Netflix Prize problem \cite{koren2009matrix}.
The basic idea is to factor the adjacency matrix of a network into the product of two matrices. The rows and columns of these matrices can be interpreted as latent features of the nodes. Latent features can be also the result of a graph embedding \cite{goyal2017graph}. In particular, there are recent attempts to apply neural networks for this purpose \cite{perozzi2014deepwalk,grover2016node2vec}. 

In this paper, we propose to use spring-electrical models to address the link prediction problem. These models have been successfully used for networks visualization \cite{fruchterman1991graph,walshaw2000multilevel,hu2005efficient}. A good network visualization usually implies that nodes similar in terms of network topology, e.g., connected and/or belonging to one cluster, tend to be visualized close to each other \cite{noack2009modularity}. Therefore, we assumed that the Euclidean distance between nodes in the obtained network layout correlates with a probability of a link between them. Thus, our idea is to use the described distance as a prediction score. We evaluate the proposed method against several popular baselines and demonstrate its flexibility by applying it to undirected, directed and bipartite networks.

The rest of the paper is organized as follows. First, we formalize the considered problem and present standard metrics for performance evaluation in Section~\ref{Problem Statement} and review related approaches which we used as baselines in Section~\ref{Baselines}.  Next, we discuss spring-electrical models and introduce our method for link prediction in Section~\ref{Spring-Electrical-Models}. 
We start a comparison of methods with a case study discussed in Section~\ref{Case-study}. Section~\ref{Results for Undirected Networks} presents experiments with undirected networks. We modify the basic model to apply it to bipartite and directed networks  in Section~\ref{Model Modifications}, followed by conclusion in Section~\ref{Conclusion}. 

\section{Problem Statement} \label{Problem Statement}
We focus on the structural definition of the link prediction problem~\cite{menon2011link}. The network with some missing edges is given and the aim is to predict these missing edges. This definition allows us to work with networks having only a single time snapshot. We also assume that there is no side information such as node or edge attributes, thus, we focus on the link prediction methods based solely on the currently observed link structure.

More formally, suppose that we have a network ${G = \langle V, E \rangle}$ without multiple edges and loops, where $V$ is the set of nodes and $E$ is the set of edges. 
We assume that $G$ is a connected graph, otherwise we change it to its largest connected component. The set of all pairs of nodes from $V$ is denoted by $U$. 

Given the network $G$, we actually do not know its missing edges. Thus, we hide a random subset of edges $E_{pos} \subset E$, while keeping the network connected. The remaining edges are denoted by $E_{train}$. Also we randomly sample unconnected pairs of nodes $E_{neg}$ from $U \backslash E$. In this way, we form ${E_{test} = E_{pos} \cup E_{neg}}$ such that ${|E_{neg}| = |E_{pos}|}$ and ${E_{test} \cap E_{train} = \emptyset}$. We train models on the network ${G' = \langle V, E_{train} \rangle}$ and try to find missing edges $E_{pos}$ in $E_{test}$.

We assume that each algorithm provides a list of scores for all pairs of nodes $(u,v) \in E_{test}$. The $score(u,v)$ characterizes similarity of nodes $u$ and $v$. The higher the $score(u,v)$, the higher probability of these nodes to be connected is. 

To measure the quality of algorithms we use the area under the receiver operating characteristic curve (AUC) \cite{hanley1982meaning}. From a probabilistic point of view AUC is the probability that a randomly selected pair of nodes from $E_{pos}$ has higher score than a randomly selected pair of nodes from $E_{neg}$:
\begin{equation*}
    \sum_{(u_p, v_p) \in  E_{pos}} 
    \sum_{(u_n, v_n) \in  E_{neg}} 
    \frac{ I \left[{score(u_p,v_p) > score(u_n,v_n)}\right] }{ | E_{pos} | \cdot | E_{neg} |},
\end{equation*}
where $I[\cdot]$ denotes an indicator function.

We repeat the evaluation several times in order to compute the mean AUC as well as the standard deviation of AUC.

\section{Related work} \label{Baselines}

In this section we describe some popular approaches to link prediction problem. The mentioned methods will be used as baselines during our experiments.

\subsection{Local Similarity Indices}

Local similarity-based methods calculate $score(u,v)$  by analyzing direct neighbours of $u$ and $v$  based on different assumptions about link formation behavior. We use $\delta(u)$ to denote the set of neighbours of $u$.

The assumption of \textit{Common Neighbours index} is that a pair of nodes has a higher probability to be connected if they share many common neighbours $$CN(u, v) := |\delta(u) \cap \delta(v)|.$$
\textit{Adamic-Adar index} is a weighted version of Common Neighbours index
$$AA(u, v) := \sum_{z \in \delta(u) \cap \delta(v)} \frac{1}{|\delta(z)|}.$$
The weight of a common neighbour is inversely proportional to its degree.

\textit{Preferential Attachment index} is motivated by Barab\'asi--Albert model~\cite{barabasi1999emergence} which assumes that the ability of a node to obtain new connections correlates with its current degree,
$$PA(u, v) := |\delta(u)| \cdot |\delta(v)|.$$

Our choice of these three local similarity indices is based on the methods comparison conducted in~\cite{liben2007link, menon2011link}. 

\subsection{Matrix Factorization}
Matrix factorization approach is extensively used for link prediction problem \cite{miller2009nonparametric,menon2010log,menon2011link,dunlavy2011temporal,acar2009link}. The adjacency matrix of the network is approximately factorized into the product of two matrices with smaller ranks. Rows and columns of these matrices can be interpreted as latent features of the nodes and the predicted score for a pair of nodes is a dot-product of corresponding latent vectors.

A truncated singular value decomposition (Truncated SVD) of matrix $A \in R^{m \times n}$ is a factorization of the form $A_r = U_r \Sigma_r V^T_r$, where $U_r \in R^{m \times r}$ has orthonormal columns, $\Sigma_r = diag(\sigma_1, ... , \sigma_r) \in R^{r \times r}$ is diagonal matrix with $\sigma_i \geq 0$ and $V_r \in R^{n \times r}$ also has orthonormal columns \cite{klema1980singular}. Actually it solves the following optimization problem:
\begin{equation*}\label{eq:svd}
\begin{aligned}
\underset{A_r: rank(A_r) \leq r}{\text{min}} \|A - U_r \Sigma_r V^T_r \|_F = \sqrt[]{\sigma_{r+1}^2 + ... + \sigma_{n}^2},
\end{aligned}
\\
\end{equation*}
where $\sigma_1, ... , \sigma_n$ are singular values of the matrix $A$.
To cope with sparse matrices we use \texttt{scipy.sparse.linalg.svds}\footnote{https://docs.scipy.org/doc/scipy-0.18.1/reference/generated/scipy.sparse.linalg.svds.html} implementation of Truncated SVD based on the implicitly restarted Arnoldi method \cite{van1996matrix}. 

Another popular approach for training latent features is a non-negative matrix factorization (NMF). NMF with $r$ components is a group of algorithms where a matrix $A \in R^{n \times m}$ is factorized into two matrices $W_r \in R^{n \times r}$ and $H_r \in R^{m \times r}$ with the property that all three matrices have non-negative elements \cite{lin2007projected}:
\begin{equation*}\label{eq:nmf}
\begin{aligned}
\underset{W_r, H_r: W_r \geq 0, H_r \geq 0}{\text{min}} \|A - W_r H_r^T \|_F.
\end{aligned}
\\
\end{equation*}

These conditions are consistent with the non-negativity of the adjacency matrix in our problem. We take as a baseline alternating non-negative least squares method with coordinate descent optimization approach \cite{cichocki2009fast} from \texttt{sklearn.decomposition.NMF}\footnote{http://scikit-learn.org/stable/modules/generated/sklearn.decomposition.NMF.html}. 

\subsection{Neural Embedding}

Several attempts to apply neural networks for graph embedding, such as DeepWalk \cite{perozzi2014deepwalk} and node2vec \cite{grover2016node2vec}, were motivated by word2vec, a widely used algorithm for extracting vector representations of words \cite{mikolov2013distributed}.  
The general idea of adopting word2vec for graph embedding is to treat nodes as ``words'' and generate ``sentences'' using random walks.  

The objective is to maximize likelihood of observed nodes co-occurrences in random walks. Probability of nodes $u$ and $v$ with latent vectors $x_u$ and $x_v$ to co-occur in a random walk is estimated using a softmax: 

\begin{equation*}\label{soft-max}
P(u | v) = \frac{\exp(x_u \cdot x_v)}{\sum_{w \in V} \exp(x_w \cdot x_v)}.
\end{equation*}

In practice a direct computation of the softmax is infeasible, thus, some approximations, such as a ``negative sampling'' or a ``hierarchical softmax'', are used  \cite{mikolov2013efficient}.

In this paper we consider node2vec which has shown a good performance in the link prediction  \cite{grover2016node2vec}.
This method generates $2^{\text{nd}}$ order random walks $c_1, \ldots, c_n$ with a transition probability defined by the following relation:
$$
P\left(c_i = x | c_{i-1} = v, c_{i-2} = t\right) \propto \begin{cases}
0,~~\text{if}~~(v,x) \notin E\\
\frac{1}{p},~~\text{else if}~~d_{tx} = 0\\
1,~~\text{else if}~~d_{tx} = 1\\
\frac{1}{q},~~\text{else if}~~d_{tx} = 2
\end{cases}
$$
where $d_{tx}$ is the graph distance between nodes $t$ and $x$. The parameters $p$ and $q$ allows to interpolate between  walks that are more akin to breadth-first or depth-first search. Generated random walks are given as input to word2vec. Finally, for each node $u$ a vector $x_u$ is assigned.

In order to estimate $score(u, v)$  we compute the dot-product of the corresponding latent vectors:

$$
node2vec(u, v) := x_u \cdot x_v.
$$

We have used a reference implementation of node2vec\footnote{https://github.com/aditya-grover/node2vec} with default parameters unless stated otherwise.

\section{Spring-Electrical Models for Link Prediction} \label{Spring-Electrical-Models}

Currently the main application of spring-electrical models to graph analisys is a graph visualization. The basic idea is to represent a graph as a mechanical system of like charges connected by springs~\cite{fruchterman1991graph}. In this system between each pair of nodes  act repulsive forces and between adjacent nodes act attractive forces. In an equilibrium state of the system, the edges tend to have uniform length (because of the spring forces), and nodes that are not connected tend to be drawn further apart (because of the electrical repulsion).

Actually, in practice edge attraction and vertex repulsion forces may be defined using functions that are not precisely based on the Hooke's and Coulomb's laws. For instance, in ~\cite{fruchterman1991graph}, the pioneering work of Fruchterman and Reingold, repulsive forces are inversely proportional to the distance and attractive forces are proportional to the square of the distance. In~\cite{walshaw2000multilevel} and~\cite{hu2005efficient} spring-electrical models were further studied and the repulsive force got new parameters $C$, $K$ and $p$, which we will discuss later. In our research, we will also use their modification with the following forces:
\begin{align}
\label{eqn:sfdp_forces}
\begin{split}
 f_r(u, v) &= -C K^{1+p}/||x_u - x_v||^p, \,\,\,\,\,\, p > 0, u \neq v; u, v \in V ,
\\
 f_a(u, v) &= ||x_u - x_v||^2/K , \,\,\,\,\,\, (u, v) \in E; u, v \in V.
\end{split}
\end{align}

\begin{figure}[H]
     \centering
     \includegraphics[width=0.27\textwidth]{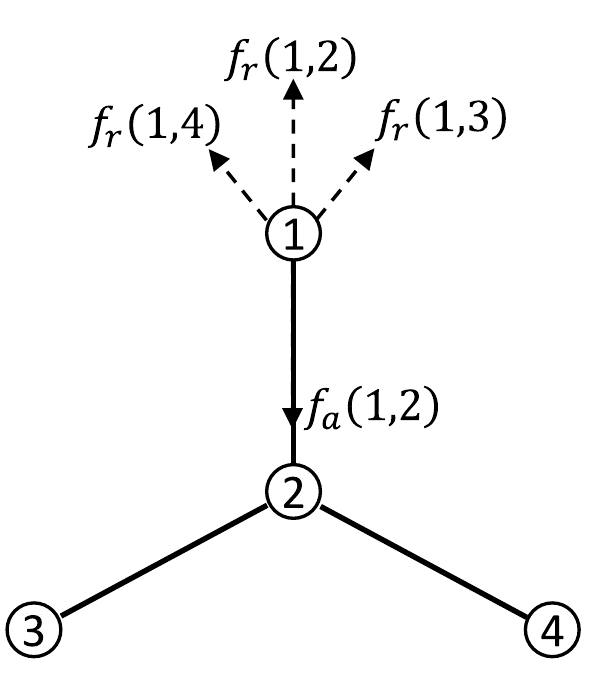}
     \caption{Spring-electrical model}%
     \label{fig:spring_electrical}%
\end{figure}

Here we denote by $\|x_u-x_v\|$ the Euclidean distance between coordinate vectors of the nodes $u$ and $v$ in a layout, and by $f_r(u, v)$ and $f_a(u, v)$ we denote values of repulsive and attractive forces, respectively. 
Figure~\ref{fig:spring_electrical} illustrates the forces acting on one of the nodes in a simple star graph.

\begin{figure*}[ht]%
    \centering
    \subfloat[SFDP]{{\includegraphics[width=0.27\linewidth,trim={3.5cm 0 2.5cm 0},clip]{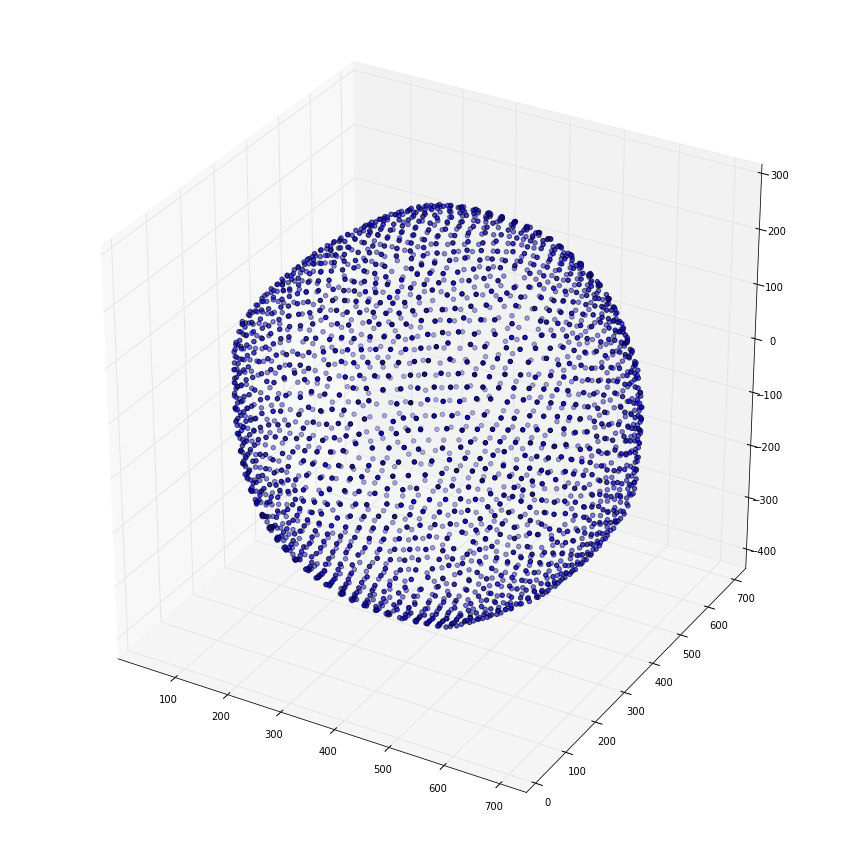} }}%
	\qquad
    \subfloat[SVD]{{\includegraphics[width=0.27\linewidth,trim={3.5cm 0 2.5cm 0},clip]{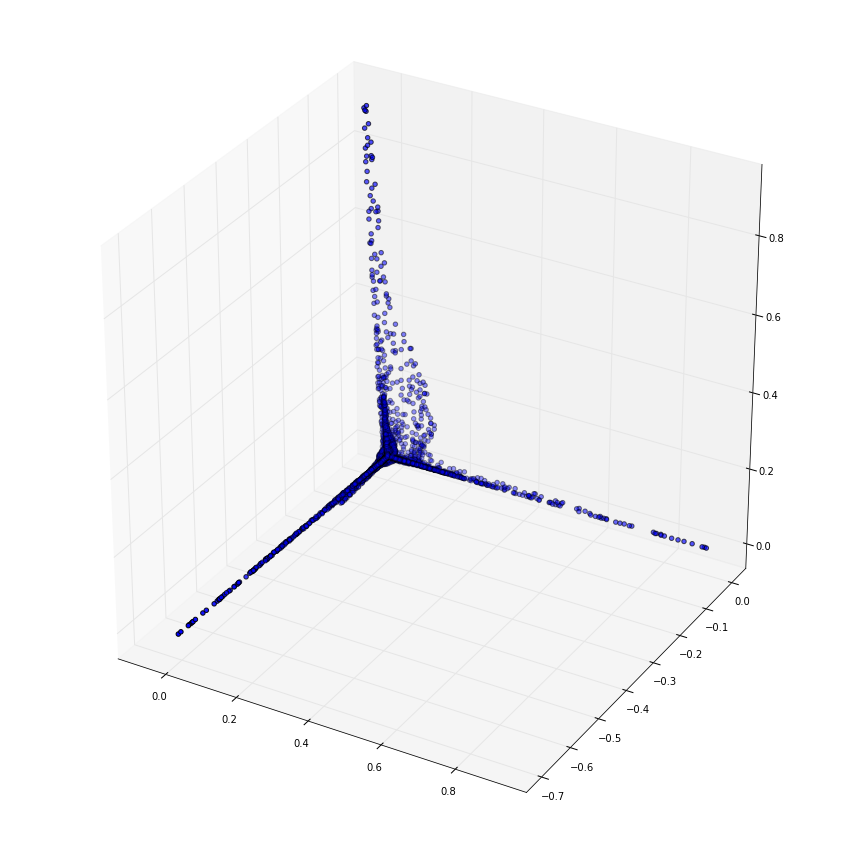} }}%
    \qquad
    \subfloat[node2vec]{{\includegraphics[width=0.27\linewidth,trim={3.5cm 0 2.5cm 0},clip]{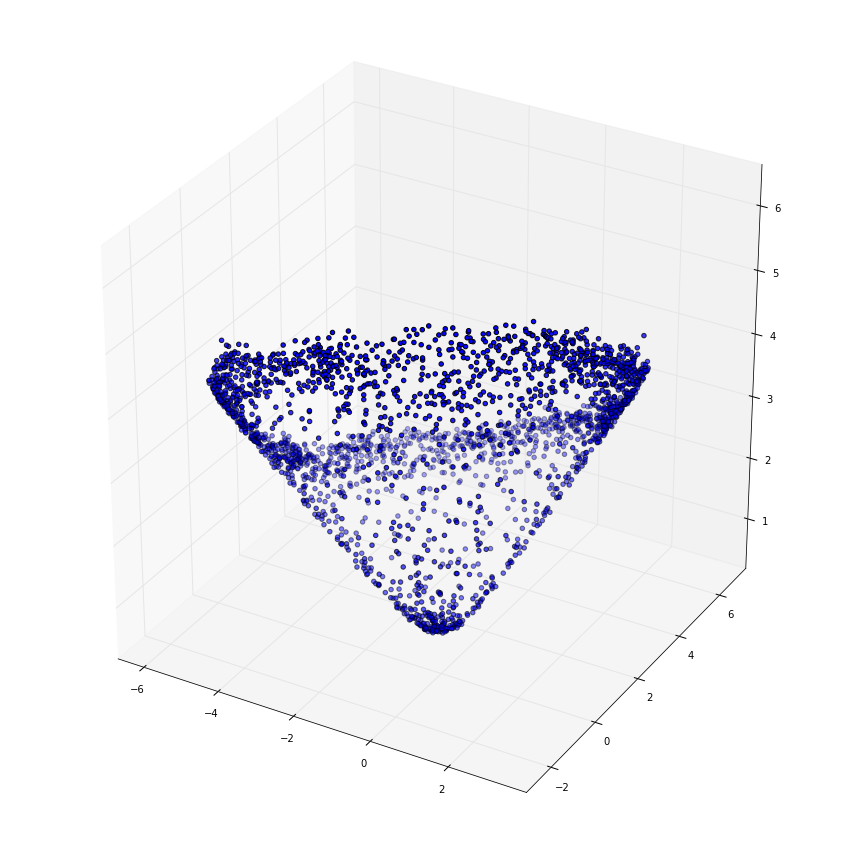} }}%
    \caption{Triangulation of 3d-sphere (3d latent features)}%
    \label{fig:sphere_viz}%
\end{figure*}

\begin{figure}[ht]
     \centering
     \includegraphics[width=0.43\textwidth]{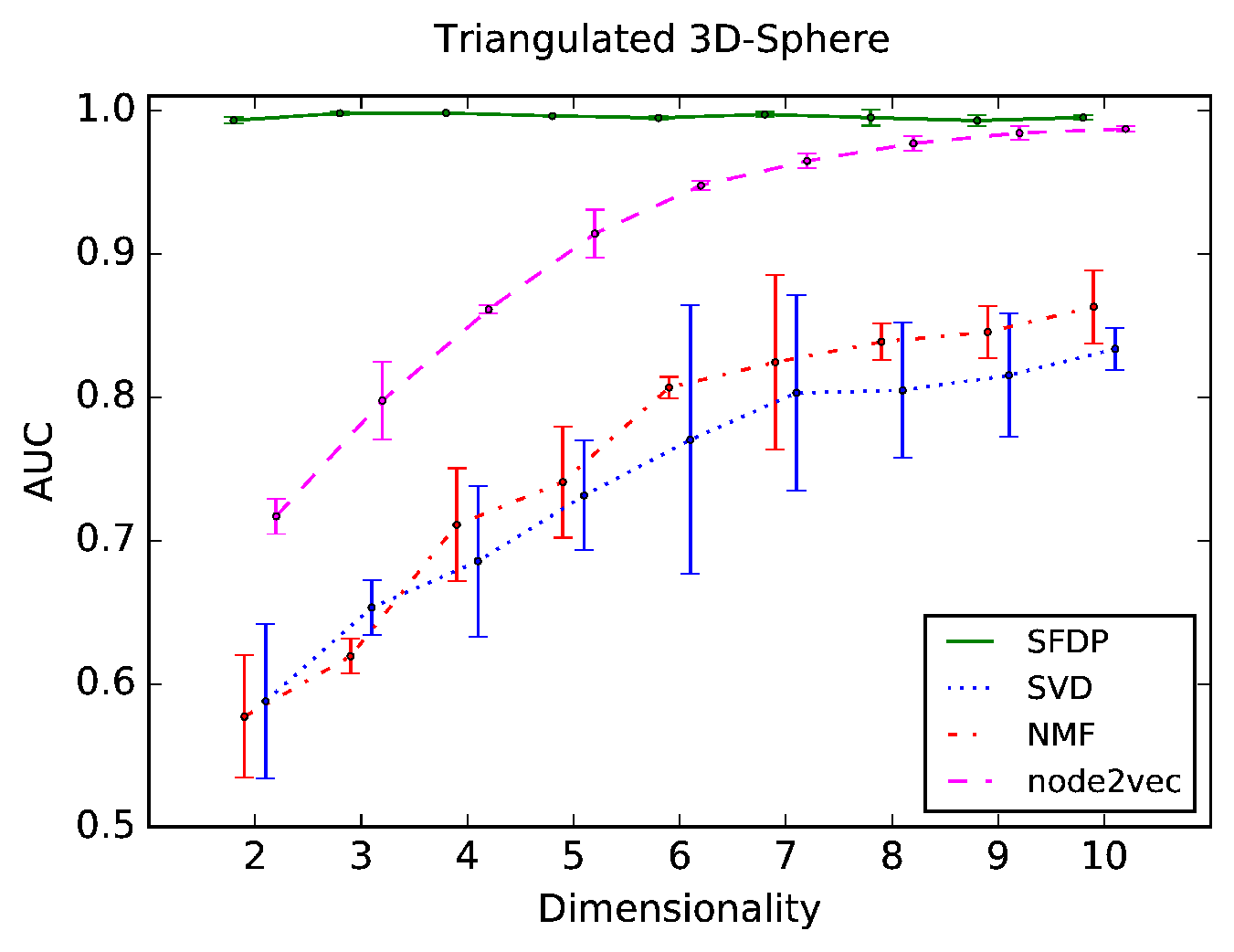}
     \caption{Triangulation of 3d-sphere (AUC scores)}%
     \label{fig:sphere_auc}%
\end{figure}

By exploiting that force is the negative gradient of energy, the force model can be transformed into an energy model, such that force equilibria correspond to (local) energy minima \cite{noack2009modularity}. An optimal layout is achieved in the equilibrium state of the system with the minimum value of the system energy. Thus, finding of an optimal layout is actually the following optimization problem:

\begin{equation}
\label{eqn:sfdp_energy}
\min_{\{x_w, w \in V \}} \left( \sum_{(u,v) \in E} \frac{||x_u - x_v||^3}{3K} + \sum_{\substack{u, v \in V\\ u \neq v}} \frac{\frac{1}{p-1} C K^{1+p} }{||x_u - x_v||^{p-1} } \right).
\end{equation}

A lot of algorithms were proposed to find the optimal layout. All of them meet two main challenges: 
(i) a computational complexity of the repulsive forces, 
(ii) a slow convergence speed and trapping in local minima.  
Note that local minima of the energy might lead to layouts with bad visualization characteristics.
We will use Scalable Force Directed Placement (SFDP) algorithm described in~\cite{hu2005efficient}, which is able to overcome both challenges. 

The  computational complexity challenge in SFDP is solved by Barnes-Hut optimization~\cite{barnes1986hierarchical}.
As a result, the total complexity of all repulsive forces calculation reduces to $O(|V|\log |V|)$, compared to a straightforward method with complexity $O(|V|^2)$. 
The second challenge is solved by a multilevel approach in combination with an adaptive cooling scheme~\cite{walshaw2000multilevel,hu2005efficient}.
The idea is to iteratively coarse the network until the network size falls below some threshold. Once the initial layout for the coarsest network is found, it is successively refined and extended to all the levels starting with the coarsest networks and ending with the original.

Let us now discuss how the model parameters influence on the equilibrium states of the system. According to Theorem~1~in~\cite{hu2005efficient}, parameters $K$ and $C$ do not change possible equilibriums and only scale them, however, they influence on the speed of convergence to an equilibrium.
As it follows from equation (\ref{eqn:sfdp_forces}), a parameter $p$ controls the strength of repulsive forces. For small values of $p$, nodes on the periphery of a layout are affected strongly by repulsive forces of central nodes. This leads to so-called ``peripheral effect'': edges at the periphery are longer than edges at the center~\cite{hu2005efficient}. On the other hand, despite larger values of $p$ reduce ``peripheral effect'', too small repulsive forces might lead to clusters collapsing \cite{noack2009modularity}. We study influence of the repulsive force exponent $p$ on the performance of our method in Section~\ref{Results for Undirected Networks}.

A good network visualization usually implies that nodes similar in terms of network topology, e.g., connected and/or belonging to one cluster, tend to be visualized close to each other~\cite{noack2009modularity}. 
Therefore, we assumed that the Euclidean distance between nodes in the obtained network layout correlates with a probability of a link between them. 
Thus, to address the link prediction problem, we first find a network layout using SFDP and then use the distance between nodes as a prediction score:
\begin{equation*}
\text{SFDP}(u, v) = \| x_u - x_v \|.
\end{equation*}

Our method can be interpreted as a latent feature approach to link prediction. 

\section{Case study} \label{Case-study}

\begin{figure*}[h]
     \centering
     \includegraphics[width=0.32\textwidth]{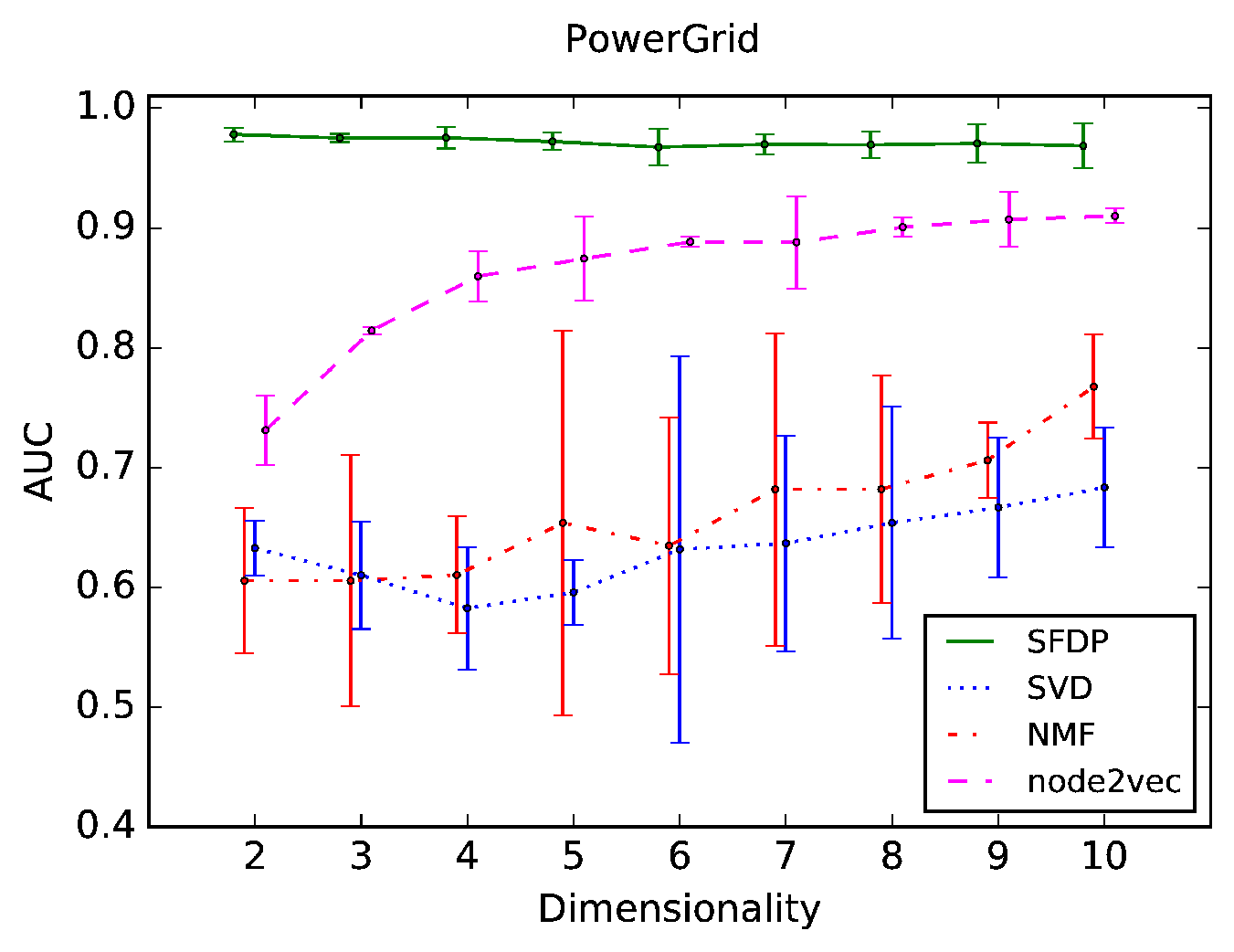} 
     \includegraphics[width=0.32\textwidth]{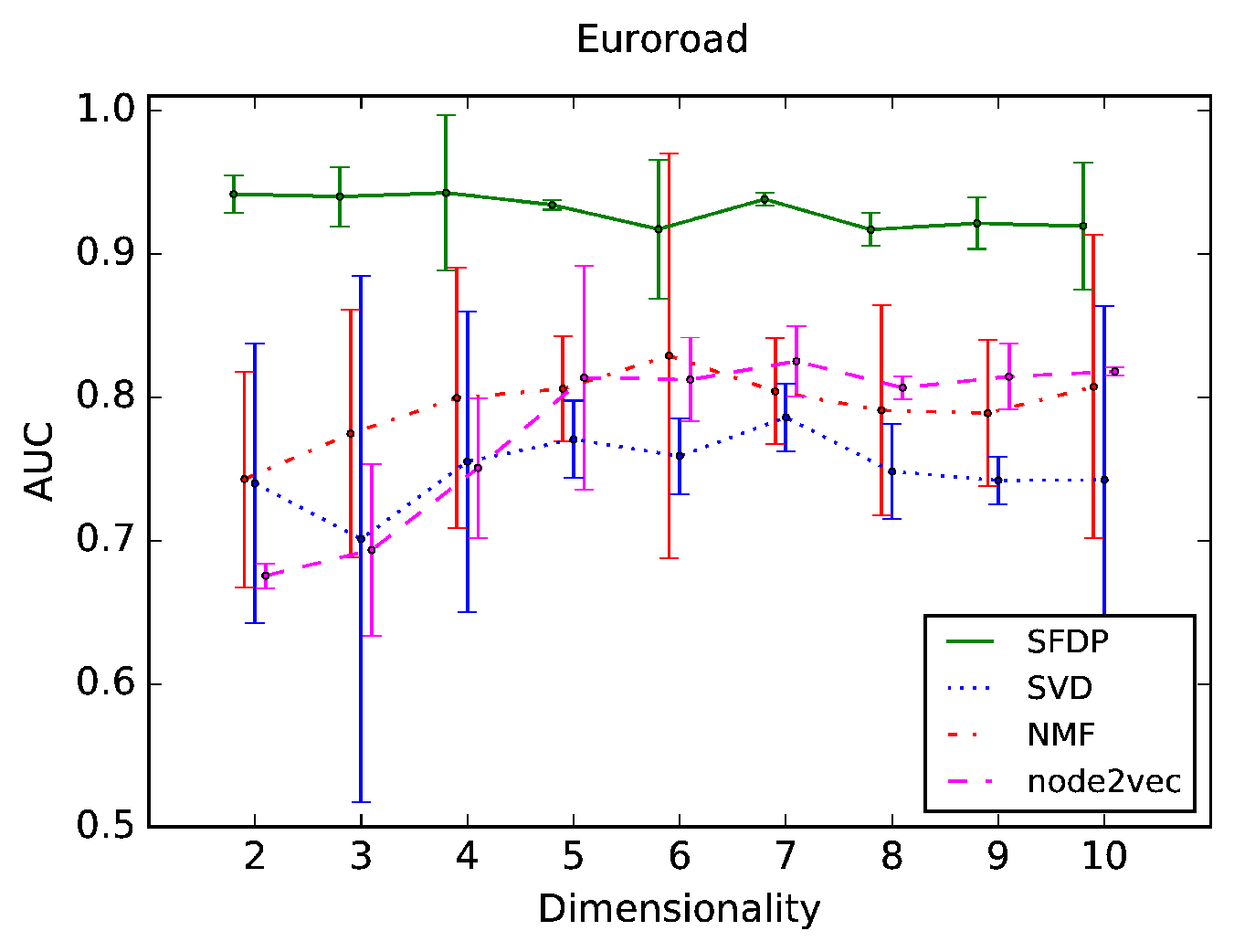} 
     \includegraphics[width=0.32\textwidth]{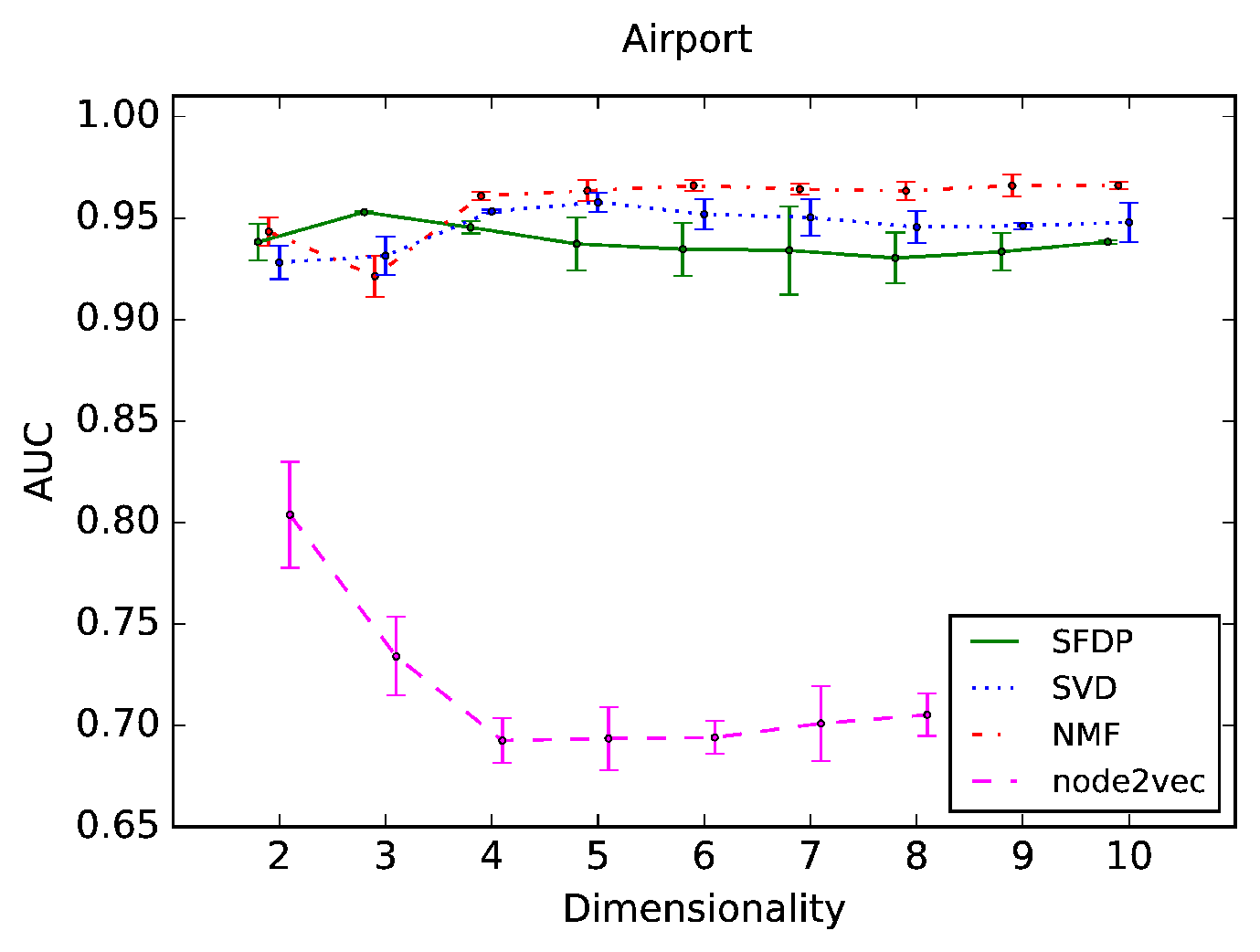} 
     
     \medskip

     \includegraphics[width=0.32\textwidth]{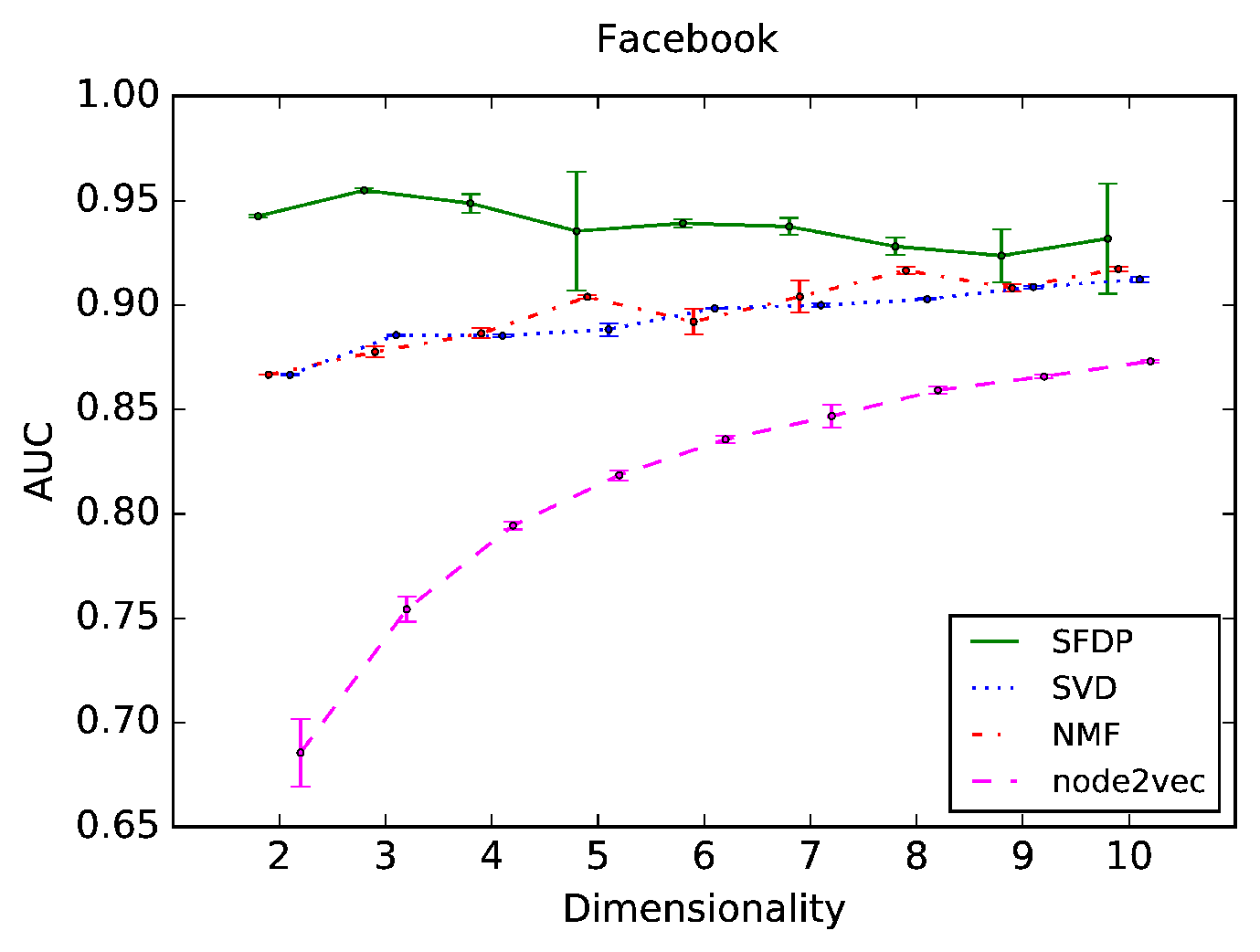} 
     \includegraphics[width=0.32\textwidth]{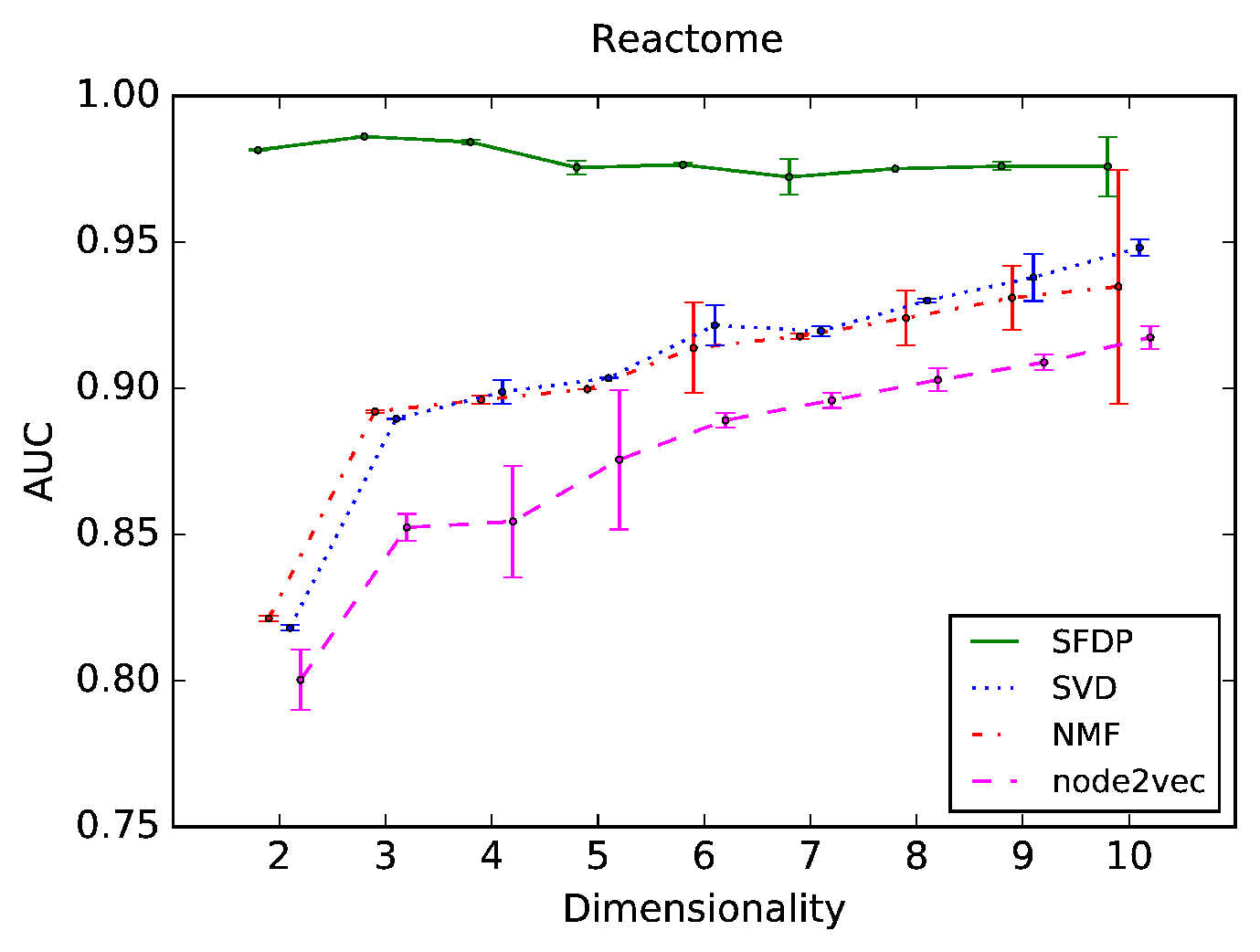} 
     \includegraphics[width=0.32\textwidth]{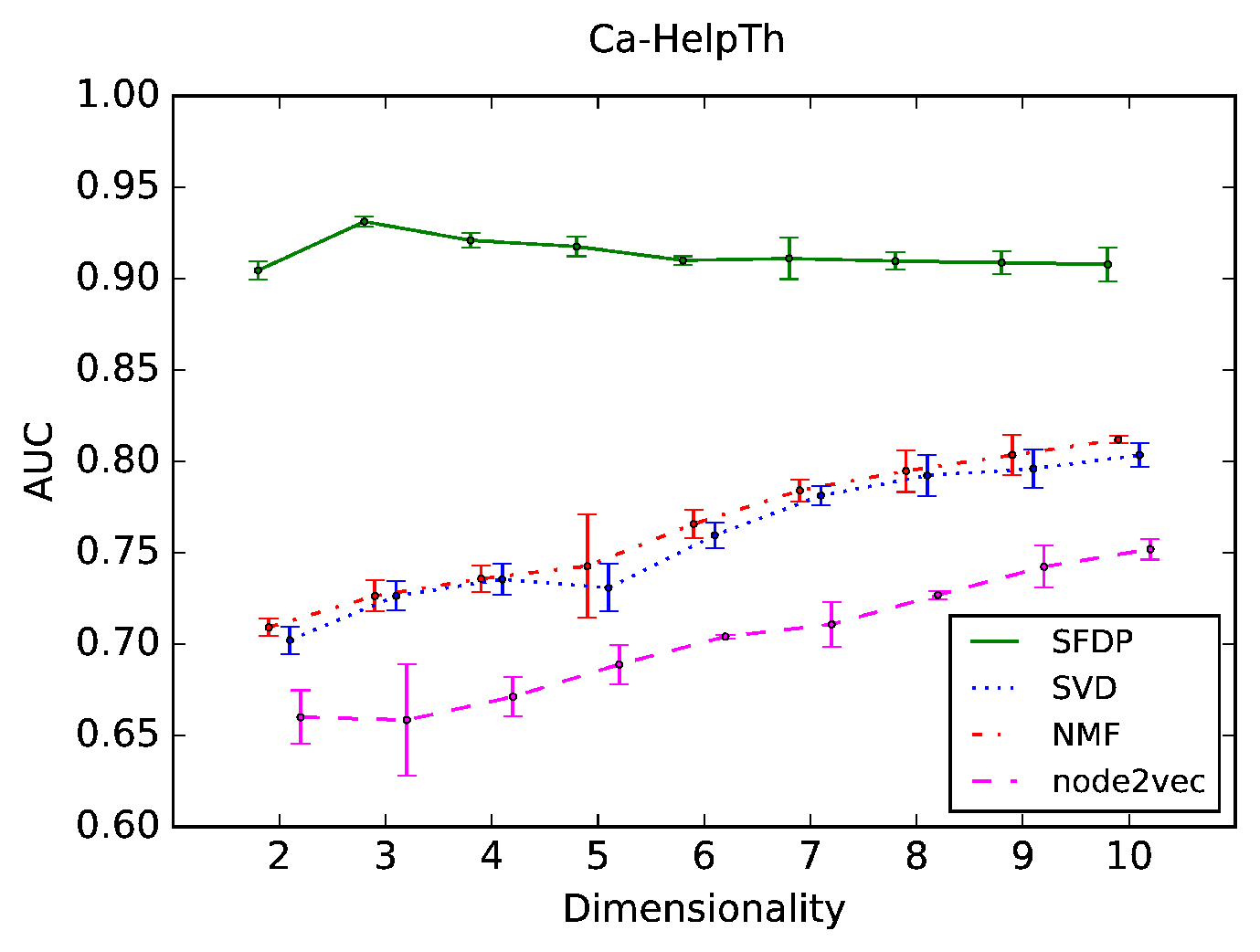} 
     \caption{Influence of dimensionality}%
     \label{fig:dimensions_1}%
\end{figure*} 

Before discussing experiments with real-world networks, we consider a graph obtained by triangulation of a three-dimensional sphere. This case study reveals an important difference between SFDP and other baselines which use latent feature space.

First, we have trained three-dimensional latent vectors  and visualized them (see Figure~\ref{fig:sphere_viz}). SFDP arranges latent vectors on the surface of a sphere as one might expect. SVD's latent vectors form three mutually perpendicular rays and node2vec places latent vectors on the surface of a cone. The reason of such a different behavior is that SFDP uses the Euclidean distance in its loss function (see equation~(\ref{eqn:sfdp_energy})), while node2vec and SVD relies on the dot-product. One can see that dot-product based methods fail to express the fact that all nodes in the considered graph  are structurally equivalent.

The difference between the dot-product and the Euclidean distance is in the way they deal with latent vectors corresponding to completely unrelated nodes. The dot-product tends to make such vectors perpendicular, while the Euclidean distance just places them ``far away''. The number of available mutually perpendicular direction is determined by the dimensionality of the latent feature space. As the result, dimensionality becomes restrictive for dot-product based methods.


In order to further support this observation we have evaluated AUC score of the discussed methods depending on the dimensionality of the latent feature space. The results are presented on Figure~\ref{fig:sphere_auc}. SFDP has good quality starting from very low dimensions. Node2vec achieves a reasonable quality in the ten-dimensional latent feature space, while SVD and NMF needs about 100 dimensions.

Our experiments with real-world networks, described above, confirm that SFDP might have a competitive quality of link prediction even in very low dimensions. This advantage might lead to some practical applications as many problems related to vector embedding are much easier in low dimensions, e.g., searching of the nearest neighbors.

     
     

\section{Experiments with Undirected Networks} \label{Results for Undirected Networks}

First, we have chosen several undirected networks in which geographical closeness correlates with the probability to obtain a  connection. Thus, the ability to infer a distance feature can be tested on them. 

\begin{itemize}
\item ``PowerGrid''~\cite{konect} is an undirected and unweighted network representing the US electric powergrid network. There are $4,941$ nodes and $6,594$ supplies in the system. It is a sparse network with average degree $2.7$. 
\item ``Euroroad''~\cite{konect}  is a road network located mostly in Europe. Nodes represent cities and an edge between two nodes denotes that they are connected by a road. This network consists of	$1,174$ vertices (cities) and $1,417$ edges (roads).
\item ``Airport''~\cite{konect} has information about $28,236$ flights between $1,574$ US airports in $2010$. Airport network has hubs, i.e. several busiest airports. Thus, the connections occur not only because of geographical closeness, but also based on the airport sizes.
\end{itemize}

We have also chosen several undirected networks of other types.

\begin{itemize}
\item ``Facebook''~\cite{konect} is a Facebook friendship network consist of
$817,035$ friendships and $63,731$ users. This network is a subset of full Facebook friendship graph.
\item ``Reactome''~\cite{konect} has information about $147,547$ interactions between $6,327$ proteins.
\item ``Ca-HepTh''~\cite{snapnets} is a collaboration network from the arXiv High Energy Physics - Theory section from January 1993 to April 2003.  The network has $9,877$ authors and $25,998$ collaborations.
\end{itemize}

All the datasets and our source code are available in our GitHub repository\footnote{https://github.com/KashinYana/link-prediction
}.
In our experiments we have used two implementations of SFDP, from \texttt{graphviz}\footnote{\url{http://www.graphviz.org}} and \texttt{graph\_tool}\footnote{\url{https://graph-tool.skewed.de/}} libraries, with the default parameters unless stated otherwise. 

{\small
\begin{table}[h]
\centering
\caption{Comparison with latent features models}\label{tab:vs_latent}
\begin{tabular}{|c||c|c|c|c|}
 \hline
Dataset    & SFDP & SVD & NMF & node2vec  \\  \hline \hline
 PowerGrid & \makecell{2d \\ \maxff{bx}{0.978}$\pm$0.005} & \makecell{30d \\ 0.848$\pm$0.007} & \makecell{40d \\  0.913$\pm$0.009} & \makecell{50d \\ \maxff{sb}{0.931}$\pm$0.011}  \\  \hline
  Euroroad & \makecell{2d \\ \maxff{bx}{0.941}$\pm$0.012} & \makecell{7d \\ 0.785$\pm$0.023} & \makecell{6d \\ 0.829 $\pm$ 0.037} & \makecell{75d \\ 0.871$\pm$0.021}  \\  \hline
 Airport & \makecell{3d \\ \maxff{sb}{0.953}$\pm$0.000} & \makecell{5d \\ \maxff{sb}{0.957}$\pm$0.005} & \makecell{6d \\ \maxff{bx}{0.966}$\pm$ 0.003} & \makecell{2d \\ 0.804$\pm$0.026}  \\  \hline \hline
 Facebook  & \makecell{3d \\ \maxf{0.951}$\pm$0.000} & \makecell{20d \\ 0.922$\pm$0.000} & \makecell{500d \\ \maxf{0.959}$\pm$0.001} & \makecell{150d \\ 0.935$\pm$0.000}  \\  \hline
 Reactome  & \makecell{3d \\ \maxff{sb}{0.986}$\pm$0.000} & \makecell{100d \\ \maxff{sb}{0.987} $\pm$0.001} & \makecell{125d \\ \maxff{bx}{0.993} $\pm$0.000} & \makecell{100d \\ 0.954$\pm$0.000}  \\  \hline
 Ca-HepTh  & \makecell{3d \\ \maxff{bx}{0.931}$\pm$0.004} & \makecell{100d \\ 0.856$\pm$ 0.005} & \makecell{150d \\ \maxff{sb}{0.921}$\pm$0.007} & \makecell{125d \\ 0.884$\pm$0.006}  \\  \hline
\end{tabular}
\end{table}
}

Following the discussion in Section~\ref{Case-study}, we have first analyzed the behavior of latent feature methods in low dimensions. On the sparse datasets ``PowerGrid'' and ``Euroroad'' we hide $10\%$ of edges in order to keep a train network connected. On other datasets $30\%$ of edges were included in a test set. We repeat this process several times and report mean AUC scores as well as 95\% confidence intervals. The results are presented on Figure~\ref{fig:dimensions_1}.

As it was expected the dot-product based methods have a clear growing trend on the most of networks, with lower performance in low dimensions. In contrast, SFDP has a good quality starting from two dimensions usually with a slight increase at dimensionality equals three and a slowly decreasing trend after this point.

We have also studied higher dimensions (up to 500 dimensions).
In Table~\ref{tab:vs_latent} for each method and dataset one can find an optimal dimension with the corresponding AUC score, standard deviation values smaller than $0.0005$ are shown as zero. Surprisingly  SFDP demonstrates competitive quality in comparison even with high-dimensional dot-product based methods. This observation suggests that real networks might have a lower inherit dimensionality than one might expect. 

\begin{figure}[h]\center{
\includegraphics[width=0.9\linewidth]{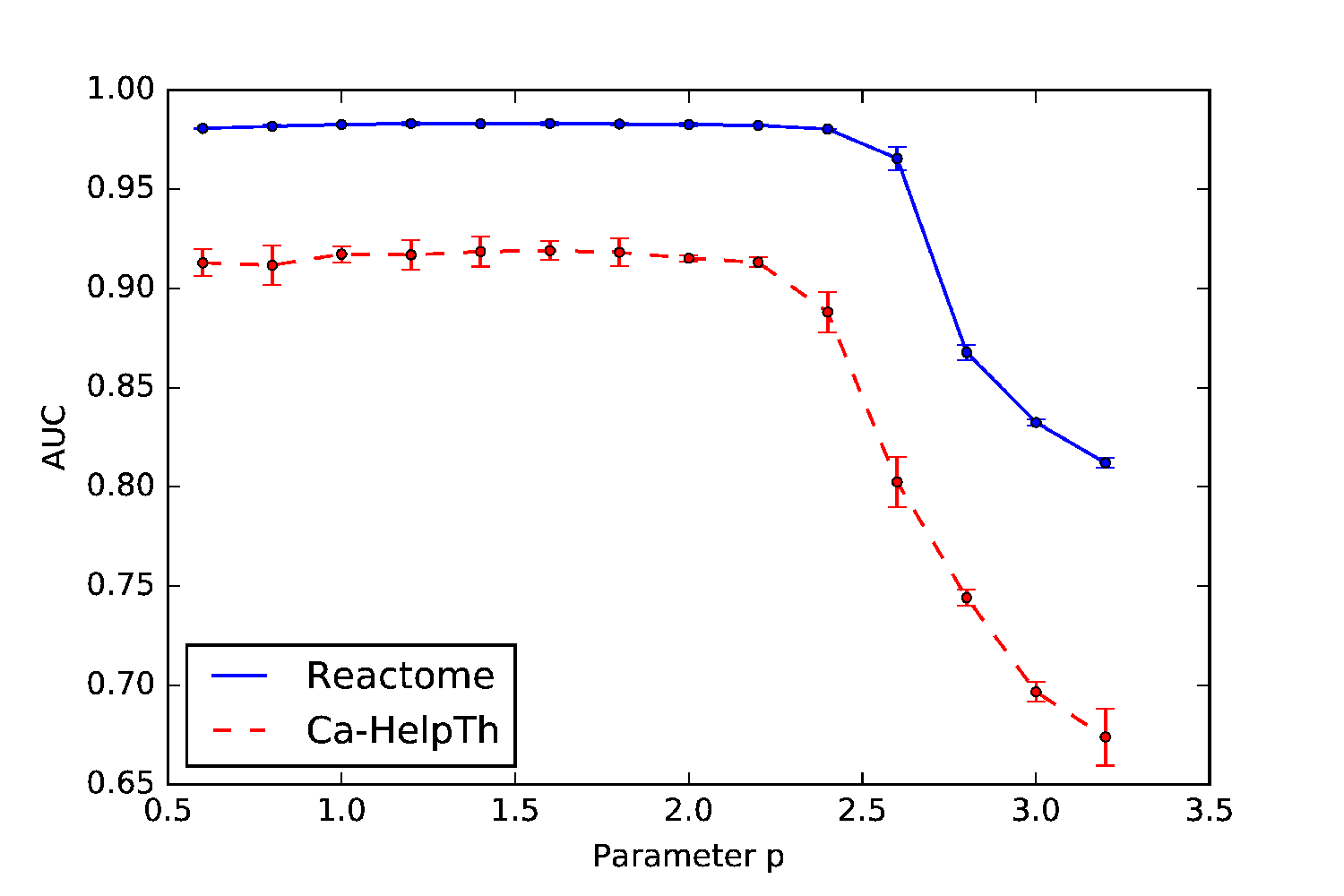}}
\caption{Influence of the repulsive force exponent}
\label{ris:p}
\end{figure}

As the influence of dimensionality on the performance of SFDP is not very significant we further focused on the two-dimensional SFDP. Speaking about other parameters, according to Section~\ref{Spring-Electrical-Models} parameters $C$ and $K$ do not change SFDP performance regarding link prediction since they only scale the optimal layout. Thus, we tried to vary the parameter $p$. Based on Figure~\ref{ris:p}, we have decided to continue using of the default value of the repulsive force exponent $p$ which equals~$2$.

{\small
\begin{table}[h]
\centering
\caption{Comparison with local similarity indices}\label{tab:2d}
\begin{tabular}{|c||c|c|c|c|}
 \hline
Dataset    & SFDP 2d & PA & CN & AA  \\  \hline \hline
PowerGrid & \makecell{\maxff{bx}{0.978}}$\pm$0.005 & 0.576$\pm$0.005 & 0.625$\pm$0.006 & 0.625$\pm$0.006  \\  \hline
Euroroad & \makecell{\maxff{bx}{0.941}$\pm$0.012} & 0.432$\pm$0.015 & 0.535$\pm$0.011 & 0.534$\pm$0.011  \\  \hline
Airport & 0.938$\pm$0.001 & 0.949 $\pm$0.000 & \makecell{\maxff{sb}{0.959}$\pm$0.000} & \makecell{\maxff{bx}{0.962}$\pm$0.001}  \\  \hline \hline
Facebook  & \makecell{\maxff{bx}{0.943}$\pm$0.000} & 0.887$\pm$0.000 & \makecell{\maxff{sb}{0.915}$\pm$0.000} & \makecell{\maxff{sb}{0.915}$\pm$0.000}  \\  \hline
Reactome  & \makecell{\maxff{sb}{0.981}$\pm$0.000} & 0.899$\pm$0.001 & \maxff{bx}{0.988}$\pm$0.000 & \makecell{\maxff{bx}{0.989}$\pm$0.000}  \\  \hline
Ca-HepTh  & \makecell{\maxff{bx}{0.905}$\pm$0.001} & 0.787$\pm$0.000 & 0.867$\pm$0.002 & 0.867$\pm$0.002  \\  \hline
\end{tabular}
\end{table}
}

Finally, we have compared SFDP with local similarity indices. The results can be found in Table~\ref{tab:2d}. SFDP has shown the highest advance on the geographical networks ``PowerGrid'' and ``Euroroad''. This observation supports our hypothesis that SFDP can infer geographical distance.
The result of SFDP on the ``Airport'' network is not so good and we link this fact to the presence of distinct hubs, we will return to this network in Section~\ref{sec:DiSFDP}. On other datasets spring-electrical approach has shown superior or competitive quality. 

\section{Model Modifications} \label{Model Modifications}

The basic spring-electrical model can be adapted to different networks types. In this section we will present possible modifications for bipartite and directed networks. 

\subsection{Bipartite Networks}\label{BiSFDP}

{
\begin{table*}[!ht]
\caption{AUC scores for bipartite datasets, |$E_{pos}$|/|$E$| = $0.3$}\label{tab:bi_SFDP}
\centering
\begin{tabular}{|c||c|c|c|c|c|c|c|c|}
 \hline
Dataset  & Bi-SFDP 2d & SFDP 2d & PA &  NMF 100d &  SVD 100d &  node2vec 100d \\ \hline \hline
 Movielens & 0.758 $\pm$0.001 & 0.755$\pm$0.005 & 
 0.773$\pm$0.000 & 
 \makecell{\maxff{sb}{0.870}$\pm$0.002} &  \makecell{\maxff{bx}{0.876}$\pm$0.000} &
0.725 $\pm$0.000 \\  \hline
 Condmat &  0.682$\pm$0.007 & \makecell{\maxff{bx}{0.910}$\pm$0.002} & 0.617$\pm$0.001 &
 0.819$\pm$0.005 & 0.787$\pm$0.004 & \makecell{\maxff{bx}{0.912} $\pm$0.001}  \\  \hline 
Frwiki & 
\makecell{\maxff{bx}{0.828}$\pm$0.005} & 
0.745$\pm$0.002        & \makecell{\maxff{sb}{0.800}$\pm$0.000} &
0.544$\pm$0.027        & 0.571$\pm$0.005 & 0.508 $\pm$	0.003 \\  \hline
\end{tabular}
\end{table*}
}

A bipartite network is a network which nodes can be divided into two disjoint sets such that edges connect nodes only from different sets. It is interesting to study this special case because the link prediction in bipartite networks is close to a collaborative filtering problem. 

We use the following bipartite datasets in our experiments:

\begin{itemize}
\item ``Movielens''~\cite{konect} dataset contains information how users rated movies on the website  \url{http://movielens.umn.edu/}. The version of the dataset which we used has $9,746$ users, $6,040$ movies and $1$ million ratings. Since we are not interested in a rating score, we assign one value to all edge weights.
\item ``Frwiki''~\cite{konect} is a network of $201,727$ edit events  done by $30,997$ users in $2,884$ articles of the French Wikipedia. 
\item ``Condmat''~\cite{konect} is an authorship network from the arXiv condensed matter section (cond-mat) from 1995 to 1999. It contains $58,595$ edges which connect $16,726$ publications and $38,741$ authors.
\end{itemize}

Let us consider the ``Movielens'' dataset. This network has two types of nodes: users and movies. When applying SFDP model to this network we expect movies of the same topic to be placed nearby. Similarly, we expect users which may rate the same movie to be located closely. In the basic spring-electrical models the repulsive forces are assigned between all nodes. It works good for visualization purpose, but it hinders formation of cluster by users and movies. Therefore, we removed repulsive forces between nodes of the same type.

Consider a bipartite network $G= \langle V, E \rangle$, which nodes are partitioned into two subsets $V = L \sqcup R$ such that $E \subset L \times R$. In our modification of SFDP model for bipartite networks, denoted Bi-SFDP, the following forces are assigned between nodes.
\begin{align}
\label{eqn:bi_sfdp_forces}
\begin{split}
 f_{r}(u, v) &= -C K^{(1+p)} / ||x_u - x_v||^p, \,\,\,\,\,\, p > 0, u \in L, v \in R ,
\\
 f_a(u, v) &= ||x_u - x_v||^2 / K , \,\,\,\,\,\, (u, v) \in E; u \in L, v \in R.
\end{split}
\end{align}
 To carry out experiments with Bi-SFDP model we have written a patch for \texttt{graph\_tools} library.

\begin{figure*}[h]%
    \centering
    \subfloat[SFPD]{{\includegraphics[width=0.35\linewidth]{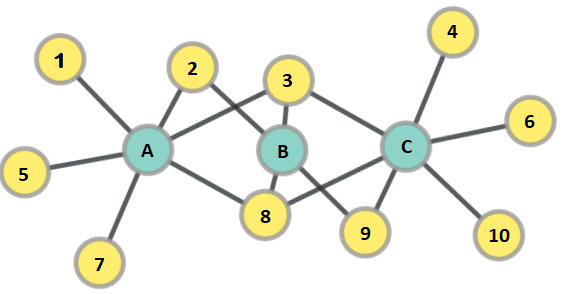}}}  \qquad   
    \subfloat[Bi-SFDP]{{\includegraphics[width=0.37\linewidth]{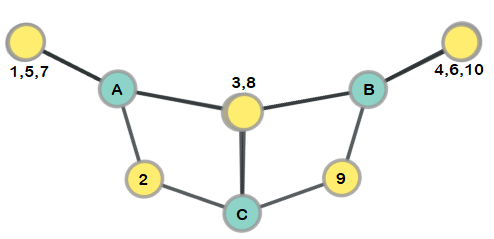} }}%
    \caption{The visualization of a bipartite network by SFDP and Bi-SFDP}%
    \label{fig:compare}%
\end{figure*}

Figure~\ref{fig:compare} demonstrates how this modification affects the optimal layout. We consider a small user--movie network of ten users and three movies. Note that on Figure~\ref{fig:compare}~(b) some of the yellow nodes were collapsed. The reason is that if we remove repulsive forces between nodes of the same type, users which link to the same movies will have the same positions. Thus, Bi-SFDP model could assigned close positions to users with the same interests.

During our preliminary experiments with bipartite networks we have found that PA demonstrates too high results. The reason is that the preferential attachment mechanism is too strong in bipartite networks. In order to focus on thinner effects governing links formation, we have changed a way to sample the set of negative pairs of nodes $E_{neg}$. A half of pairs of nodes in $E_{neg}$ were sampled with probability proportional to the product of its degrees, such pairs of nodes are counterexamples for the preferential attachment mechanism.

 The results of the experiments are summarized in Table~\ref{tab:bi_SFDP}. Baselines CN and AA are not included in the table, because their scores are always equal to zero on bipartite networks.

Bi-SFDP modification has shown the increase in quality compared with the basic SFDP model on ``Movielens'' and  ``Frwiki''  datasets. It means that our assumption works for these networks. In contrast, on the ``Condmat'' standard SFDP and node2vec outperforms all other baselines. In general, the results for bipartite graphs are very dataset-dependent.

\subsection{Directed Networks}\label{sec:DiSFDP}

Spring-electrical models are not suitable for predicting links in directed networks because of the symmetry of the forces and the distance function. Therefore, we first propose to transform the original directed network.

\begin {figure}
\centering
\subfloat[][Graph $G$]{
\begin{tikzpicture}[->,>=stealth',shorten >=1pt,auto,node distance=3cm,
                     thick,main node/.style={circle,draw,font=\sffamily\Large\bfseries}]                   
   \node[main node] (1) {$A$};
   \node[main node] (2) [below left of=1] {$B$};
   \node[main node] (3) [below right of=1] {$C$};
   \path[]
      (1) edge [] node[left] {} (2)
      (2) edge [] node[left] {} (1)
      (2) edge [] node[left] {} (3)
      (3) edge [] node[right] {} (1);
\end{tikzpicture}
}
\hspace{1cm}
\subfloat[][Graph $G'$]{
\begin{tikzpicture}[->,>=stealth',shorten >=1pt,auto,node distance=2cm,
                     thick,main node/.style={circle,draw,font=\sffamily\Large\bfseries}]
   \node[main node] (1) {$A_{out}$};
   \node[main node] (2) [right of=1] {$B_{out}$};
   \node[main node] (3) [right of=2] {$C_{out}$};
   \node[main node] (4) [below of=1] {$A_{in}$};
   \node[main node] (5) [below of=2] {$B_{in}$};
   \node[main node] (6) [below of=3] {$C_{in}$};   
   \path[]
      (1) edge [] node[left] {} (5)
      (2) edge [] node[left] {} (4)
      (2) edge [] node[left] {} (6)
      (3) edge [] node[left] {} (4);
\end{tikzpicture}
}
\caption{Di-SFDP graph transformation}%
\label{fig_disfdp}%
\end{figure}
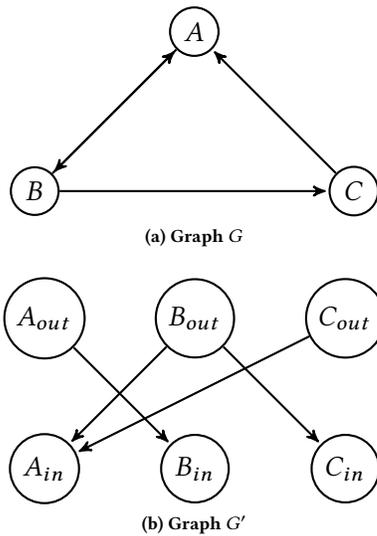

Given a directed network $G= \langle V, E \rangle$, we obtain an undirected bipartite network $G'=  \langle V', E'  \rangle$, $V' = L \sqcup R$, $E' \subset L \times R$ by the following process. Each node $u \in V$ corresponds to two nodes $u_{out} \in L$ and $u_{in} \in R$. One of the nodes is responsible for outgoing connections, the other one for incoming connections. Thus, for each directed edge $(u, v) \in E$ an edge $(u_{out}, v_{in})$ is added in $E'$.  Figure~\ref{fig_disfdp} illustrates the described transformation. As the result, $G'$ has information about the all directed edges of $G$.

Then Bi-SFDP model can be applied to find a layout of the network $G'$. Finally, prediction scores for pairs of nodes from the network $G$ can be easily inherited from the layout of the network $G'$.
We have called this approach Di-SFDP and have tested on the following datasets.
\begin{itemize}
\item ``Twitter''~\cite{konect} is a user-user network, where directed edges represent the fact that one user follows the other user. The network contains $23,370$ users and $33,101$ follows.
\item ``Google+''~\cite{konect} is also a user-user network. Directed links indicate that one user has the other user in his circles. There are $23,628$ users and $39,242$ friendships in the network. 
\item ``Cit-HepTh''~\cite{konect} has information about  $352,807$ citations among $27,770$ publications in the arXiv High Energy Physics Theory (hep-th) section. 
\end{itemize}

All pairs of nodes $(u, v)$ such that $(v, u) \in E$ but $(u, v)\not \in E$ we call \textit{difficult pairs}. They can not be correctly scored by the basic SFDP model. It is especially interesting to validate models on such pairs of nodes. Therefore, in our experiments a half of pairs of nodes in $E_\text{neg}$ are difficult pairs.

The experiment results are shown in Table~\ref{tab:di_SFDP}. The baselines PA, CC and AA can be also calculated on $G'$, but their quality is close to a random predictor.
One can see that Di-SFDP outperforms other baselines on two of the datasets and have a competitive quality on the last one. Note that out-of-box node2vec can not correctly score difficult pairs of nodes as it infers only one latent vector for each node, while other methods has two latent vectors, one is responsible for outgoing connections and another one for incoming connections.

Di-SFDP has also helped us to improve quality on the ``Airport'' dataset. Despite ``Airport'' is an undirected network, due to presence of hubs it has a natural orientation of edges. Thus, our idea was to first orient edges from the nodes of low degrees to the nodes of high degrees and then apply Di-SFDP. This trick allowed us to improve the mean AUC from 0.938 to 0.972.

{\small
\begin{table}[H]
\caption{AUC scores for directed datasets, |$E_{pos}$|/|$E$| = $0.3$}\label{tab:di_SFDP}
\centering
\begin{tabular}{|c||c|c|c|c|}
\hline
Dataset  & Di-SFDP 2d &  NMF 100d & SVD 100d & node2vec 100d \\  \hline \hline
Twitter & \makecell{\maxff{bx}{0.952} $\pm$0.002} & 0.783$\pm$0.010 & 0.694$\pm$0.014 
& 0.550 $\pm$ 0.001
\\  \hline
Google+ & \makecell{\maxff{bx}{0.998} $\pm$0.000} & \makecell{\maxff{sb}{0.936}$\pm$0.007} & 0.466$\pm$0.033 
& 0.449 $\pm$ 0.002
\\  \hline
Cit-HepTh & \makecell{\maxff{sb}{0.836} $\pm$0.003} & \makecell{\maxff{bx}{0.842}$\pm$0.002} & \makecell{\maxff{sb}{0.838}$\pm$0.001}
& 0.679$\pm$0.000 
\\  \hline
\end{tabular}
\newline
\vspace*{2mm}
\newline
\begin{tabular}{|c||c|c|}
\hline
Dataset  & Di-SFDP 2d &  SFDP 2d \\  \hline \hline
Airport & \makecell{\maxff{bx}{0.972} $\pm$0.001} & 0.938$\pm$0.001 \\  \hline
\end{tabular}
\end{table}
}

\section{Conclusion} \label{Conclusion}
In this paper we proposed to use spring-electrical models to address the link prediction problem. We first applied the basic SFDP model to the link prediction in undirected networks and then adapted it to bipartite and directed  networks by introducing two novel methods, Bi-SFDP and Di-SFDP. The considered models demonstrates superior or competitive performance of our approach over several popular baselines. 

A distinctive feature of the proposed method in comparison with other latent feature models is a good performance even in very low dimensions. This advantage might lead to some practical applications as many problems related to vector embedding are much easier in low dimensions, e.g., searching of the nearest neighbors. 
On the other hand this observation suggests that real networks might have lower inherit dimensionality than one might expect.

We consider this work as a good motivation towards a new set of research directions. 
Future research can be focused on choosing an optimal distance measure for latent feature models and deeper analysis of inherit networks dimensionality.



\end{document}